\documentclass[conference]{IEEEtran}
\usepackage{graphicx}
\usepackage{color}
\usepackage{multirow}

\usepackage{flushend}

\usepackage{cite}

\usepackage{amsmath}

\usepackage{comment}

\ifCLASSINFOpdf
\else
\fi
\usepackage{array}

\def\qed{\hbox{${\vcenter{\vbox{
        \hrule height 0.4pt\hbox{\vrule width 0.4pt height 6pt
        \kern5pt\vrule width 0.4pt}\hrule height 0.4pt}}}$}}
 
\newcolumntype{C}{>{\centering\arraybackslash}X}
\newcolumntype{L}{>{\raggedright\arraybackslash}X}
\newcolumntype{R}{>{\raggedleft\arraybackslash}X}
\begin{document}
\title{Diverse Adaptive Bulk Search: a Framework for Solving QUBO Problems on Multiple GPUs}

\author{
\IEEEauthorblockN{Koji Nakano, Daisuke Takafuji,  and Yasuaki Ito}
\IEEEauthorblockA{Graduate School of Advanced Science and Engineering, Hiroshima University\\
Kagamiyama 1-4-1, Higashihiroshima, 739-8527 Japan}
\\
\IEEEauthorblockN{Takashi Yazane, Junko Yano, Shiro Ozaki, Ryota Katsuki, and Rie Mori}
 \IEEEauthorblockA{Research and Development Headquarters, NTT DATA Corporation\\
Toyosu Center Bldg, Annex, 3-9, Toyosu 3-chome, Koto-ku, Tokyo 135-8671, Japan}
}



\maketitle


\begin{abstract}
Quadratic Unconstrained Binary Optimization (QUBO)  is a combinatorial optimization to find an optimal binary solution vector that minimizes the energy value defined by a quadratic formula of binary variables in the vector.
As many NP-hard problems can be reduced to QUBO problems, considerable research has gone into developing QUBO solvers running on various computing platforms such as quantum devices, ASICs, FPGAs, GPUs, and optical fibers.
This paper presents a framework called Diverse Adaptive Bulk Search (DABS), which has the potential to find optimal solutions of many types of QUBO problems.
Our DABS solver employs a genetic algorithm-based search algorithm featuring three diverse strategies: multiple search algorithms, multiple genetic operations, and multiple solution pools.
During the execution of the solver, search algorithms and genetic operations that succeeded in finding good solutions are automatically selected to obtain better solutions.
Moreover, search algorithms traverse between different solution pools to find good solutions.
We have implemented our DABS solver to run on multiple GPUs.
Experimental evaluations using eight NVIDIA A100 GPUs confirm that our DABS solver succeeds in finding optimal or potentially optimal solutions for three types of QUBO problems.
\end{abstract}


\section{Introduction}


\subsection{Background: Ising model and QUBO model}
\begin{figure}
\centering
\includegraphics{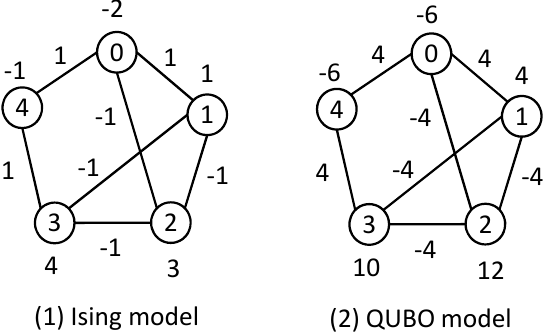}
\caption{Examples of BQMs, an Ising model and a QUBO model}
\label{fig:ising}
\end{figure}

\emph{Quantum annealing}~\cite{Kadowaki98} is a metaheuristic to find an optimal spin vector
of an Ising model by a process using quantum fluctuations.
\emph{An Ising model}~\cite{Brush67} is defined by a weighted undirected graph $G=(V,E)$ with a set of $n$ nodes $V=\{0, 1, \ldots, n-1\}$
and edge set $E$.
Each undirected edge $(i,j)\in E$ is assigned a weight $J_{i,j}$ called \emph{an interaction}, and each node $i$ is assigned
a weight $h_i$ called \emph{a bias}.
\emph{The Hamiltonian} $H(S)$ of a \emph{spin} (or \emph{qubit}) vector $S=s_0s_1\cdots s_{n-1}$ ($s_i\in \{-1,+1\}$ for all $i$)
is defined by the following formula:
\begin{align}
H(S) & =  \sum_{(i,j)\in E} J_{i,j}s_is_j+\sum_{i\in V}h_is_i.
\end{align}
Fig.~\ref{fig:ising}~(1) shows an example of an Ising model.  
Quantum annealing aims to search for an optimal spin vector with the smallest Hamiltonian over all spin vectors.
For example, Ising model in Fig.~\ref{fig:ising}~(1) has an optimal solution $S=[+1,-1,-1,-1,+1]$ with the smallest Hamiltonian $H(S)=-14$.
D-Wave Systems developed a quantum annealer called D-Wave 2000Q~\cite{McGeoch19}
with a 2048-node Chimera graph~\cite{Vert19} quantum network topology.
The interactions and biases of the quantum annealer are programmable, and its quantum annealing
searches for an optimal solution of the corresponding Ising model.
Usually, quantum annealing is repeated hundreds or thousands of times, and the best solution is output.
Later, D-Wave Systems released a larger quantum annealer called D-Wave Advantage~\cite{Advantage},
which can handle Ising models with a 5760-node Pegasus graph~\cite{DWaveAdvantage19}.
Although D-Wave quantum annealers work for particular graph topologies,
they can handle Ising models with different graph topologies by embedding them in Chimera/Pegasus graphs.
For example, a 177-node complete graph can be embedded into a Pegasus graph; hence,
D-Wave Advantage can be used to perform quantum annealing for 177-spin Ising models with any graph topology.

This paper mainly focuses on \emph{Quadratic Unconstrained Binary Optimization (QUBO) models}.
A QUBO model is defined by a weighted undirected graph $G=(V,E)$ with $n$ nodes, as shown in Fig.~\ref{fig:ising}~(2).
Let $W_{i,j}$ ($(i,j)\in E$) denote the weight of an edge connecting nodes $i$ and $j$,
and $W_{i,i}$ ($0\leq i\leq n-1$) denote the weight of node $i$.
A QUBO problem for a given model $W=(W_{i,j})$ aims to find a binary vector $X=x_0x_1\cdots x_{n-1}$ ($x_i\in \{0,1\}$ for all $i$)
that minimizes \emph{the energy} defined as follows:
\begin{align}
E(X) &=  \sum_{(i,j)\in E\cup V\times V} W_{i,j}x_ix_j.
\label{eq:QUBO}
\end{align}
Any Ising model can equivalently be converted to a QUBO model with the same graph topology, and vice versa~\cite{Tanahashi19,Tao20},
such that spin values $-1/+1$ correspond to binary values $0/1$, respectively.
Fig.~\ref{fig:ising}~(1) and~(2) are examples of an Ising model and QUBO model equivalent to each other.
Since $E(X)=H(S)+6$ always holds for all corresponding $S$ and $X$, we have an optimal solution $X=[1,0,0,0,1]$ with energy $E(X)=-8$.
Thus, any Ising problem solver including D-Wave quantum annealers can be used as QUBO solvers and vice versa.
Hence, Ising models and QUBO models are collectively called \emph{Binary Quadratic Models (BQMs)}.
Further, we consider solvers for BQMs with Chimera/Pegasus topologies as simulators
for D-Wave quantum annealers\cite{Imanaga21}.
Such simulators provide insights into the quantum supremacy of quantum annealers.
As long as efficient simulators running on non-quantum devices exist, quantum supremacy cannot be achieved.

Because optimization problems for BQMs are NP-hard, it is not possible to design a polynomial time algorithm
using classical computers with digital circuit devices of polynomial size as long as $P\neq NP$.
If ideal quantum annealers based on quantum mechanics that can find optimal solutions for large BQMs were available,
many NP-hard problems could be solved in some quantum annealing time.
However, current quantum annealers are not sufficiently powerful for such a purpose.
The number of bits is too small and the probability of finding optimal solutions by quantum annealing is insufficient due to undesirable flux noise~\cite{Zaborniak21}.
Hence, as an alternative to an ideal quantum annealer, BQM solvers
on various non-quantum computing platforms, such as ASICs~\cite{Oku19,Yamamoto21}, 
FPGAs~\cite{Kagawa21,Goto21},  
GPUs~\cite{Okuyama19,Yasudo-JPDC22}, and optimal fibers~\cite{Inagaki16,Honjo21} have been proposed.
Further, D-Wave Systems released a hybrid BQM solver~\cite{D-Wave-Hybrid20} that
uses both a classical computer and quantum annealer
to find solutions for large BQMs with up to 1,000,000-node complete graphs.

\subsection{Our contributions}
\begin{figure}
\centering
\includegraphics{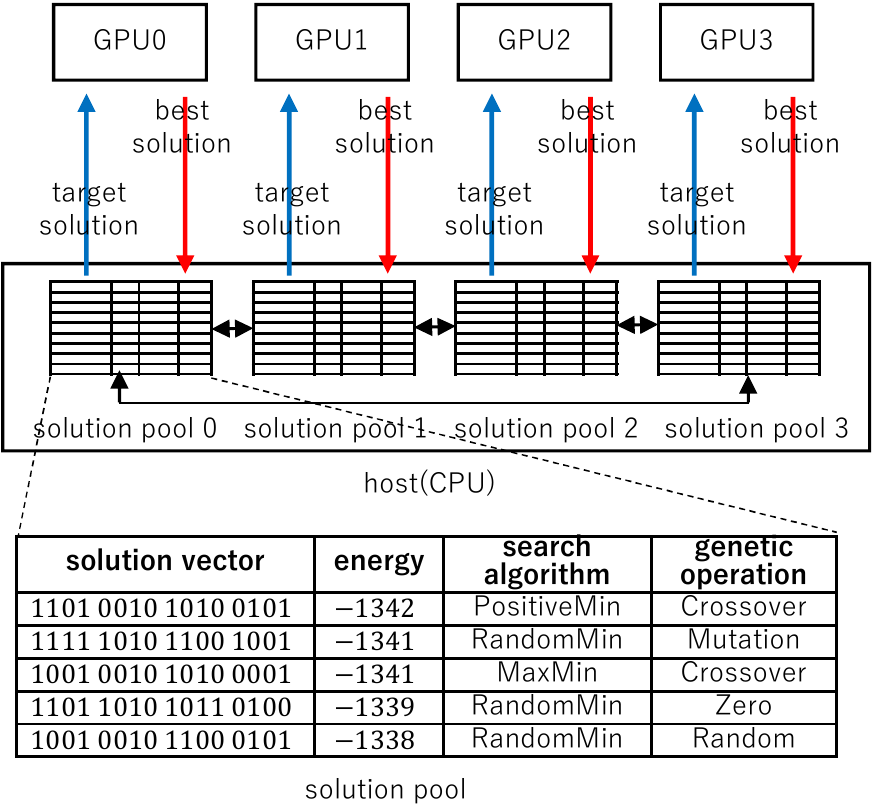}
\caption{Illustrating the architecture of our DABS solver: GPUs executes search algorithms from given target solutions and host CPU performs genetic operations on solution pools to generate target solutions.}
\label{fig:outline}
\end{figure}

The main contribution of this paper is to present a framework called the \emph{Diverse Adaptive Bulk Search (DABS)} for solving QUBO models
and to implement it on multiple GPUs.
It has various diverse features:
(1) multiple search algorithms, (2) multiple genetic operations, and (3) multiple solution pools.
Fig.~\ref{fig:outline} illustrates the architecture of our DABS solver.
The host has multiple solution pools, each associated with a GPU, and storing good solutions obtained by the GPU
with energy values, search algorithms, and genetic operations.
An Open MP thread is assigned to a solution pool, and it generates target solution vectors
by genetic operations, such as mutation and crossover for selected solutions in the solution pool.
The generated target solution vectors are sent to the associated GPU, in which multiple CUDA blocks with multiple threads work in parallel.
Each CUDA block executes a search algorithm specified by the host starting from the target solution vector.
When the search terminates, the best solution vector obtained during the execution is sent to the host.
An Open MP thread running on the host receives and inserts it in the solution pool if it is better than the worst solution vector in the solution pool.


The well-known \emph{No Free Lunch Theorem (NFLT)}~\cite{nofreelunch17} implies that there exists no heuristic search algorithm
that can solve all types of QUBO problems efficiently.
A heuristic search algorithm that works efficiently for a particular QUBO model may not work well for other models.
Moreover, we might have to select different heuristic algorithms during the execution.
For example, assume that we have two heuristic algorithms for a particular QUBO model:
Heuristic algorithm $A$ can find a good solution very quickly but has little chance to find an optimal solution, while
heuristic algorithm $B$ runs very slowly but has a good chance to find an optimal solution from a good non-optimal solution.
For solving this QUBO model, we may first execute $A$.
After finding a good solution by $A$, we execute $B$ for finding an optimal solution from a good solution.
Our DABS solver is designed so that $A$ or $B$ are selected appropriately during the execution.
For the diversity of local search algorithms,
we implemented five local search algorithms: MaxMin, CyclicMin, RandomMin, PositiveMin, and TwoNeighbor.
All of them repeat bit flipping of a solution vector and traverse the $n$-bit search space of QUBO models.
During the execution of the DABS solver, a local search algorithm that succeeded in finding better solutions is executed 
more frequently than the others.
In addition, a genetic operation from which better solutions have been found is executed more frequently.
We implemented eight genetic operations: Mutation, Crossover, Xrossover, Zero, One, IntervalZero, Best, and Random.
Our DABS solver dynamically selects suitable search algorithms/genetic operations during execution.

A genetic algorithm (GA) using multiple solution pools is known as the island model~\cite{Whitley99}.
During the execution of a GA, a solution pool tends to be filled with relatives of the best solution in the pool.
Because target solution vectors obtained by genetic operations for such a solution pool may be similar,
search algorithms may traverse the same narrow region of the $n$-bit search space, degrading the search performance.
The island model using multiple solution pools may improve the solution diversity.
We introduce genetic operation Xrossover, which is a crossover operation between solution pools.
In the Xrossover genetic operation, a target solution is generated by the crossover operation of two solution vectors, one from the current solution pool
and the other from a neighbor solution pool in Fig.~\ref{fig:outline}.
Thus, a local search for this target solution will traverse the $n$-bit search space between solution pools,
and the search area will be expanded.

In~\cite{Yasudo-JPDC22}, we have presented a GA-based QUBO solver called the Adaptive Bulk Search (ABS) that executes the CyclicMin search algorithm only.
Moreover, a single genetic operation, mutation after crossover, was performed.
Because the search algorithm and genetic operation are fixed, 
once it stacks non-optimal local minimum solutions,
it may not be possible to leave for a better local minimum.
Hence, this paper significantly extends the ABS solver with a variety of diverse features to improve the search performance for various QUBO problems.


We evaluated the performance of our DABS solver for QUBO models reduced from the MaxCut problem,
Quadratic Assignment Problem (QAP), and Quantum Annealer Simulation Problem (QASP).
For comparison, we used Gurobi 9.5.1~\cite{Gurobi}, a Mixed-Integer Programming (MIP) solver supporting quadratic formulas.
We also used a quantum annealer D-Wave Advantage~\cite{Advantage} and D-Wave hybrid solver~\cite{D-Wave-Hybrid20}.
The experimental results show that our QUBO solver succeeded in finding optimal or potentially optimal solutions for all QUBO models.
We define a \emph{potentially optimal solution} as a solution with circumstantial evidence of optimality as follows:
(1) our DABS solver can always find it very quickly, and no better solution can be found in a number of repeated trials with a much longer computing time,
(2) Gurobi optimizer fails to find a better solution even if it is given as a start solution, and
(3) no previously published work has found better solutions.
The experimental results also show that Gurobi optimizer and D-Wave Advantage, and D-Wave Hybrid solvers cannot find potentially optimal solutions for most QUBO models.

\subsection{Organization}
The remainder of this paper is organized as follows.
For readers who are not familiar with QUBO models,
we first show how combinatorial optimization problems can be reduced to QUBO models
in Section~\ref{sec:QUBO}.
The problems include MaxCut, QAP, and QASP.
In Section~\ref{sec:algorithms}, we present the search algorithms used in our DABS solver
for traversing an $n$-bit search space of QUBO models on a GPU:
Greedy, Straight, MaxMin, CyclicMin, RandomMin, PositiveMin, and TwoNeighbor.
We also explain the batch search, which repeatedly executes search algorithms on the GPU.
Section~\ref{sec:GA} explains the GA-based algorithm to generate target solutions for our DABS solver.
In Section~\ref{sec:GPU}, we show how we implement our DABS solver to run on multiple GPUs.
We show experimental results for QUBO problems in Section~\ref{sec:experiment}.
Section~\ref{sec:concl} concludes the paper.

\section{Optimization problems and reduction to QUBO models}
\label{sec:QUBO}
This section shows that three optimization problems, the MaxCut, QAP,
and QASP can be reduced to QUBO models.

\subsection{MaxCut problem}
Suppose we have a weighted undirected graph with $n$ nodes 0, 1, $\ldots$ $n-1$.
Let $w_{i,j}$ denote the weight of edge connecting nodes $i$ and $j$.
MaxCut aims to find a node separation into a subset $S$ and complement $\overline{S}$ such that
the total weight of edges connecting two separated nodes, one in $S$ and the other in $\overline{S}$, is maximized.
A graph of any MaxCut problem with $n$ nodes can be reduced to a QUBO model with $n$ bits
such that each node $i$ ($0\leq i\leq n-1$) is assigned to a bit $i$ and it is in $S$ if and only if $x_i=1$.
For the reduction, a quadratic formula $w_{i,j}(2x_ix_j-x_i^2-x_j^2)$ is generated
for each edge with weight $w_{i,j}$.
Clearly, the formula takes $-w_{i,j}$ if $x_i\neq x_j$ and 0 otherwise.
Thus, by combining all formulas, we have a quadratic formula of a QUBO model
such that an optimal solution with the minimum energy $E(X)$ corresponds to that of the MaxCut problem
with the maximum cut value $-E(X)$.

\subsection{Quadratic assignment problem (QAP)}
\begin{figure}[!ht]
\centering
\includegraphics{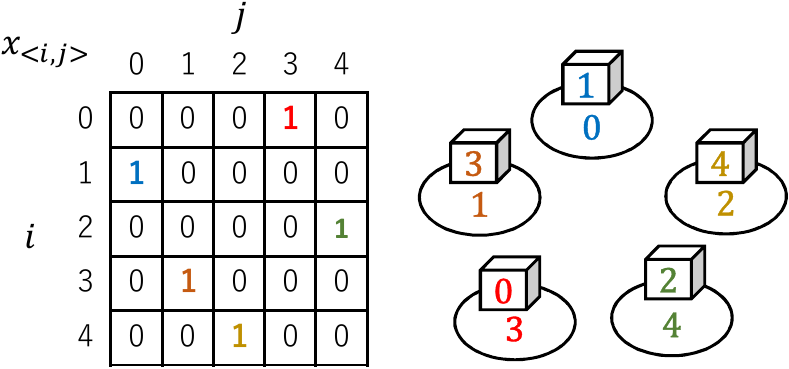}
\caption{Example of a solution vector with a one-hot encoding}
\label{fig:one-hot}
\end{figure}

Suppose that, for two sets of $n$ facilities $F$ and $n$ locations $L$,
we are given the logistic flow $l(i,j)$ between facilities $i$ and $j$ ($0\leq i,j\leq n-1$) and the distance $d(i,j)$ between locations $i$ and $j$ ($0\leq i,j\leq n-1$).
A one-to-one mapping $g:\{0,1,\ldots,n-1\}\rightarrow \{0,1,\ldots,n-1\}$ from $F$ to $L$
is used to determine the location of the facility such that each facility $i$ ($0\leq i\leq n-1$) is placed at a location $g(i)$.
The QAP aims to find $g$ that minimizes the total cost $C(g)=\sum l(i,j)\cdot d(g(i),g(j))$.
We can reduce this problem to a QUBO model with $N=n^2$ bits by \emph{a one-hot encoding}~\cite{Glover18,Tosun22}.
Let $\langle i,j\rangle=i\cdot n+j$, and
let $x_{\langle i,j\rangle}$ ($0\leq i,j\leq N-1$) denote an $N$-bit vector $X$ of the QUBO model,
which represents a one-to-one mapping such that $x_{\langle i,j\rangle}=1$ if and only if $j=g(i)$.
Thus, $x_{\langle i,j\rangle}$ must be a one-hot encoding such that every row and every column of $x_{\langle i,j\rangle}$ has exactly one 1 entry,
as shown in Fig.~\ref{fig:one-hot}.
For example, since facility 0 is placed at location 3, $g(0)=3$ and $x_{\langle 0,3\rangle}=1$ hold.
If this condition of one-hot encoding is satisfied, $X$ is said to be \emph{a feasible solution}. 
A QUBO model $W_{\langle i,j\rangle,\langle i',j'\rangle}$ corresponding to a QAP can be obtained as follows:
\begin{align*}
W_{\langle i,j\rangle,\langle i',j'\rangle} &= l(i,i')\cdot d(j,j')\quad\mbox{if $i\neq i'$ and $j\neq j'$,}\\
 & = -p \quad \mbox{if $i=i'$ and $j=j'$ and}\\
 & = p \quad \mbox{if  ($i\neq i'$ and $j=j'$) or ($i= i'$ and $j\neq j'$),}
\end{align*}
where $p$ is a large constant called the \emph{penalty term}.
The readers should have no difficulty confirming that
$E(X)=C(g_X)-np$ always holds for any feasible solution $X$,
where $g_X$ denotes the one-to-one mapping $g$ corresponding to $X$.
If $X$ is not a feasible solution, $E(X)\geq -(n-1)p$ always holds.
Thus, by selecting a large penalty $p$ appropriately
for an optimal solution $X$ of the corresponding QUBO model $W_{\langle i,j\rangle,\langle i',j'\rangle}$
gives an optimal solution $g_X$ of the original QAP problem.

The QAP is harder than the Traveling Salesperson Problem (TSP)
because the TSP can be solved by a QAP algorithm by setting a circular logistic flow of the facilities.
Moreover, QUBO models reduced from QAPs are difficult for most QUBO solvers,
because $n!$ feasible solutions are local minima, and local search algorithms must perform hill climbing or jumping to go from one local minimum
to another.

\subsection{Quantum Annealer Simulation Problem (QASP)}
Recall that D-Wave Advantage~\cite{Advantage} is a quantum annealer that searches for an optimal solution of an Ising model with
a 5760-node Pegasus graph topology.
Owing to faulty nodes and edges of the quantum network, D-Wave Advantage 4.1 can handle Ising models of a subgraph with 5,627 nodes and 40,279 edges.
The interactions $J_{i,j}$ and biases $h_i$ of the Ising model can take values in the ranges $[-1.0,+1.0]$ and  $[-4.0,+4.0]$, respectively.
\emph{The resolution} of an Ising model is an integer number $r$ ($\geq1$) such that the values of all interactions and biases are multiples of $1\over r$.
For example, an Ising model with resolution 2 can take interactions  $-1, -0.5,  0, +0.5, +1$ and biases $-4, -3.5, -3, \ldots, +3, +3.5, +4$.
The API library provided by D-Wave Systems automatically scales
interactions and biases of an Ising model so that the values fit in the ranges $[-1.0,+1.0]$ and  $[-4.0,+4.0]$.
Thus, we can think that an Ising model with resolution $r$ can take integer interactions and biases in ranges $[-r,+r]$ and $[-4r,+4r]$, respectively.
Because a quantum annealer operates interactions and biases as analog values, higher resolution values may not be managed accurately.
Hence, the resolution should be as low as possible for quantum annealers to obtain good solutions for Ising models.
Recall that any Ising model can be converted to a QUBO model with the same topology~\cite{Tanahashi19,Tao20}.
For benchmarking, we generated a random Ising model with a resolution of the real D-Wave Advantage graph such that
values of $J_{i,j}$s and $h_i$s are non-zero and chosen uniformly at random.
For example, if the resolution is 2, then the value of each $J_{i,j}$ takes an integer value from $-2$, $-1$, $+1$, or $+2$, with equal probability.
Because D-Wave Advantage is sensitive to the resolution, a benchmark QUBO model for the real D-Wave Advantage simulation should
be generated from an Ising model with a specified resolution.

\section{Search algorithms for QUBO models}
\label{sec:algorithms}
For later reference, let $\sigma: \{0,1\}\rightarrow \{-1, +1\}$ denote a one-to-one mapping such that
$\sigma(x)=2x-1$.
Clearly, $\sigma(0)=-1$ and $\sigma(1)=1$.
Let $f_i(X)$ denote the $n$-bit vector obtained by flipping bit $i$:
\begin{align*}
f_i(X)&=x_0x_1\cdots x_{i-1}\overline{x_i}x_{i+1}\cdots x_{n-1},
\end{align*}
where $\overline{x_i}=1-x_i$.
Further, let 
\begin{align*}
\Delta_i(X)&=E(f_i(X))-E(X).
\end{align*}
denote the energy value gained by a flipping bit $i$ of $X$.

\subsection{Incremental search algorithms}
The energy $E(X)$ of a given solution vector $X$ can be computed directly by calculating Eq. (\ref{eq:QUBO}) with $O(n^2)$ computational cost.
As this is too expensive, we should avoid this direct energy computation.
In this subsection, we explain \emph{the incremental search algorithm} for QUBO models, 
which is a family of local search algorithms maintaining the current solution vector $X$, energy $E(X)$,
and $\Delta_k(X)$ for all $k$ ($0\leq k\leq n-1$).
The incremental search algorithm does not calculate Eq. (\ref{eq:QUBO}) to obtain energy $E(X)$.
Initially, $X=(x_i)$ is a zero vector, that is, $x_i=0$ for all $i$.
Thus, $E(X)=0$ and $\Delta_k(X)=W_{k,k}$ for all $k$ from Eq. (\ref{eq:QUBO}).
Basically,
an incremental search algorithm determines a bit to be flipped
from the values of $\Delta_0(X), \Delta_1(X), \ldots, \Delta_{n-1}(X)$.
For example, we can select a bit $i$ with the minimum $\Delta_i(X)$.
We flip the selected bit $i$, that is, $X\leftarrow f_i(X)$ is performed.
We then update $E(X)$ and $\Delta_k(X)$ for all $k$ ($0\leq k\leq n-1$).

We will show that this update operation can be done with only $O(n)$ computational cost.
Since we have the values of $E(X)$ and $\Delta_i(X)$,
the value of $E(f_i(X))=E(X)+\Delta_i(X)$ can be updated with $O(1)$ computational cost for any $i$.
We will show that for any $k$ and $i$, $\Delta_k(f_i(X))-\Delta_k(X)$ can be computed with $O(1)$ computational cost.
From Eq. (\ref{eq:QUBO}), the value of $\Delta_k(X)$ can be computed by the following formula:
\begin{align}
\lefteqn{\Delta_k(X)=E(f_k(X))-E(X)}\notag\\
&=  \sum_{0\leq j<k}W_{j,k}x_j(\overline{x_k}-x_k)+\sum_{k< j<n}W_{k,j}(\overline{x_k}-x_k)x_j\notag\\
&  \qquad  +W_{k,k}(\overline{x_k}^2-x_k^2)\notag\\
&= -\sum_{0\leq j<n}W_{j,k}x_j\sigma({x_k})+W_{k,k}\overline{x_k}.  \label{Delta_k}
\end{align}
From Eq. (\ref{Delta_k}),
for all $i \neq k$, we have,
\begin{align}
\lefteqn{\Delta_k(f_i(X))-\Delta_k(X)} \notag\\
&= -W_{i,k}(\overline{x_i}-x_i)\sigma(x_k) = W_{i,k}\sigma(x_i)\sigma(x_k).
\label{ineqk}
\end{align}
Moreover, if $i=k$, we have,
\begin{align}
\lefteqn{\Delta_k(f_k(X))=E(f_k(f_k(X))-E(f_k(X)) }\notag\\
&= E(X)-E(f_k(X)) = -\Delta_k(X). \label{jeqk}
\end{align}
Hence, if we have $\Delta_k(X)$, we can compute $\Delta_k(f_i(X))$ with $O(1)$ computational cost.
Therefore, after a bit is flipped, $E(X)$ and $\Delta_k(X)$ can be updated for all $k$ ($0\leq k\leq n-1$) with $O(n)$ computational cost.

The incremental search algorithm repeats bit flipping operations to traverse the $n$-bit vector search space
to search for an optimal or good solution.
We extend the incremental search algorithm to record the best solution.
For this purpose, we added two variables {\it BEST} and $E({\it BEST})$, which store the best $n$-bit vector obtained so far and its energy value,
respectively.
The incremental search algorithm searches all 1-bit neighbors of $X$, and if a better solution than ${\it BEST}$ is found,
${\it BEST}$ and $E({\it BEST})$ are updated.
The steps of the incremental search algorithm are summarized as follows:

\noindent [Incremental search algorithm]\\
{\bf Step~1} Compute $E(f_i(X))=\Delta_i(X)+E(X)$ for all $i$ ($0\leq i\leq n-1$).
Compute ${\it minE}=\min\{E(f_i(X))\mid 0\leq i\leq n-1\}$, and  let $j$ be an index satisfying
$\Delta_j(X)={\it minE}$.
If $E({\it BEST})>E(f_j(X))$, then we update ${\it BEST}$ and $E({\it BEST})$ by $f_j(X)$ and $E(f_i(X))$, respectively.\\
{\bf Step~2}
Determine a bit $i$ to be flipped from the values of $\Delta_0(X), \Delta_1(X), \ldots, \Delta_{n-1}(X)$ and/or
the other values.\\
{\bf Step~3} 
Update $X$ and $E(X)$ by $f_i(X)$ and $\Delta_i(X)+E(X) $.\\
Moreover, every $\Delta_k(X)$ is updated by Eq.~(\ref{ineqk}) or~(\ref{jeqk}).\\
{\bf Step~4}
If a termination condition is not satisfied, go to Step~1.
Otherwise, output ${\it BEST}$ and $E({\it BEST})$ and then terminate. \\

Step~1 searches all 1-bit neighbors $f_0(X)$, $f_1(X)$, $\ldots$, $f_{n-1}(X)$ of $X$, and if a solution better than
the current best exists, ${\it BEST}$ is updated by it.
We can design a particular search algorithm as an incremental search algorithm.
The followings are such search algorithms.

\subsubsection{Greedy search algorithm}
Step~2 finds index $i$ with minimum $\Delta_i(X)$.
In Step~4, the algorithm terminates if $\Delta_k(X)\geq 0$ holds for all $k$.
Hence, when the algorithm terminates,
$\Delta_k(X)\geq 0$ holds for all $k$, that is, $X$ is a local minimum solution.

\subsubsection{Straight search algorithm}
A target solution vector $D=(d_i)$ is given, and the current solution vector $X=(x_i)$ moves toward $D$ as follows.
We select a bit $i$ with minimum $\Delta_i(X)$ over all bits satisfying $d_i\neq x_i$.
Clearly, this operation decreases the Hamming distance of $X$ and $D$ by one,
hence $X$ approaches $D$.
This algorithm terminates when $X=D$.

\subsubsection{MaxMin search algorithm}
Steps~1--4 are repeated for a predetermined fixed number $T$ of iterations.
Let $t$ ($1\leq t\leq T$) be a loop-counter representing the number of executed iterations.
We determine a threshold value $d$ for $\Delta_i(X)$ and randomly select a bit $i$ satisfying $\Delta_i(X)\leq d$.
A threshold value $d$ is determined as follows.
Let ${\it min\Delta}=\min\{\Delta_i(X) \mid 0\leq i\leq n-1\}$ and ${\it max\Delta}=\max\{\Delta_i(X) \mid 0\leq i\leq n-1\}$.
Further, let $D(t)=(1-({T-t\over T})^3)\cdot{\it min\Delta}+({T-t\over T})^3\cdot{\it max\Delta}$.
Clearly, $D(t)$ is a decreasing function taking a value between ${\it min\Delta}$ and ${\it max\Delta}$.
In each $t$-th iteration, a threshold value $d$ is selected from $[{\it min\Delta},D(t)]$ uniformly at random.
Step~2 randomly selects a bit $i$ satisfying $\Delta_i(X)\leq d$.
Since $d$ is not less than ${\it min\Delta}$, such a bit always exists.
The behavior of the MaxMin search algorithm is similar to that of simulated annealing~\cite{Laarhoven87},
because $D(t)$ is a decreasing function and a bit $i$ with large $\Delta_i(X)$ is selected with smaller probability
in later iterations.

Note that the operation performed in each $t$-th iteration depends on the value of $t$.
For later reference, we call such an algorithm \emph{an iteration-dependent algorithm}.

\subsubsection{CyclicMin search algorithm}
This algorithm is an iteration-dependent algorithm.
Let $w(t)=\max((t/T)^3n,c)$, where $c$ is a small constant number less than $n$, say $c=32$.
Clearly $w(t)$ is an increasing function taking values in $[c,n]$.
Suppose that $n$ bits of a QUBO model are placed cyclically and a window slides on the $n$-bit circle.
First, a window of width $w(0)$ is placed from bit 0 to $w(0)-1$.
A bit $i$ with minimum $\Delta_i(X)$ over all bits in the window is selected and flipped.
Next, a window of width $w(1)$ slides and is placed from bit $w(0)$ to $w(0)+w(1)-1$.
Again, a minimum bit in the window is selected and flipped.
The same operation of the sliding window is performed repeatedly.
More specifically, in each $t$-th iteration ($1\leq t\leq T$), a bit $i$ with minimum $\Delta_i(X)$ in the window of size $w(t)$ is selected and flipped.

Note that this search algorithm does not use random numbers.
Moreover, in each iteration, only $\Delta_i(X)$ of a bit $i$ in a window is read. 
Thus, this search algorithm can be implemented to run very efficiently on a GPU~\cite{Yasudo-JPDC22}.
Clearly, because the window size is larger in later iterations, a bit $i$ with large $\Delta_i(X)$ is selected with a smaller probability in later iterations.
Thus, the behavior of the CyclicMin algorithm is similar to that of simulated annealing.

\subsubsection{RandomMin search algorithm}
This algorithm is an iteration-dependent algorithm.
We use a probability function $p(t)=\max({(t/T)}^3,c)$, which is an increasing function of $t$, where $c$ is a small constant probability, say $32\over n$.
We select each bit $i$ as a candidate bit with probability $p(t)$.
Hence, expected $np(t)$ bits are selected.
Subsequently, a bit $i$ with minimum $\Delta_i(X)$ over all candidate bits is selected and flipped.
Because later iterations have more candidate bits, a bit $i$ with large $\Delta_i(X)$ is selected with smaller probability in later iterations.
Thus, the behavior of the RandomMin search algorithm is similar to that of simulated annealing.

\subsubsection{PositiveMin search algorithm}
Let ${\it posmin}\Delta(X)=\min\{\Delta_i(X)\mid \Delta_i(X)> 0\}$ be the minimum of $\Delta_i(X)$ with positive value.
Each bit $i$ is selected as a candidate bit if $\Delta_i(X)\leq {\it posmin\Delta(X)}$.
We select a bit from the candidate bits at random and flip it.
The PositiveMin search algorithm has been presented and implemented in an FPGA QUBO solver~\cite{Kagawa21}.
If the current solution $X$ is close to a local minimum, the number of candidate bits is small, and
a bit $i$ with $\Delta_i(X)={\it posmin}\Delta(X)>0$ is selected with a higher probability.
This fact helps to go from one local minimum to another.

\subsubsection{TwoNeighbor search algorithm}
This algorithm traverses all 1-bit neighbors of $X$ in $2n-1$ iterations (or flips).
Recall that Step~1 of the incremental search algorithm can find a better solution than the current {\it BEST} if it is a 1-bit neighbor of $X$.
The TwoNeighbor search algorithm is designed so that it searches all 2-bit neighbors of $X$.
For this purpose, Step~2 flips bits 0, 1, 0, 2, 1, 3, 2, 4, 3, 5,  $\ldots$ in turn to traverse all 1-bit neighbours.
For example,  suppose that $n=6$ and $X=000000$.
All 1-bit neighbors of $X$ are $100000$, $010000$, $001000$, $000100$, $000010$, and $000001$.
In $2n-1=11$ flip operations, $X$ stores values as follows:
\begin{align*}
&000000 \stackrel{0}{\rightarrow} {\bf 100000}\stackrel{1}{\rightarrow} 110000\stackrel{0}{\rightarrow} {\bf 010000}\stackrel{2}{\rightarrow}\\
&011000 \stackrel{1}{\rightarrow} {\bf 001000} \stackrel{3}{\rightarrow} 001100\stackrel{2}{\rightarrow} {\bf 000100} \stackrel{4}{\rightarrow}\\
&000110 \stackrel{3}{\rightarrow} {\bf 000010}\stackrel{5}{\rightarrow} 000011\stackrel{4}{\rightarrow} {\bf 000001}.
\end{align*}
Since all 1-bit neighbors are traversed and Step~1 of the incremental search algorithm searches all 1-bit neighbors of current $X$,
it can search all 2-bit neighbors of $X$.
In addition, as some of the 2-bit neighbors, such as $110000$, are traversed, it partially searches 3-bit neighbors.


For later reference, we call the MaxMin, CyclicMin, RandomMin, PositiveMin, and TwoNeighbor search algorithms \emph{the main search algorithms}.

\subsubsection{Tabu search}
We can apply \emph{the tabu search} technique~\cite{Glover97}
to MaxMin, CyclicMin, RandomMin, and PositiveMin search algorithms.
A tabu period $t$ is specified in the tabu search.
If a bit is flipped, we do not flip it again in the next $t$ iterations.
It can be flipped again after $t$ iterations.
By the tabu search technique, we can avoid stacking a particular local minimum solution.

\subsection{Batch search}
\begin{figure}[!ht]
\centering
\includegraphics{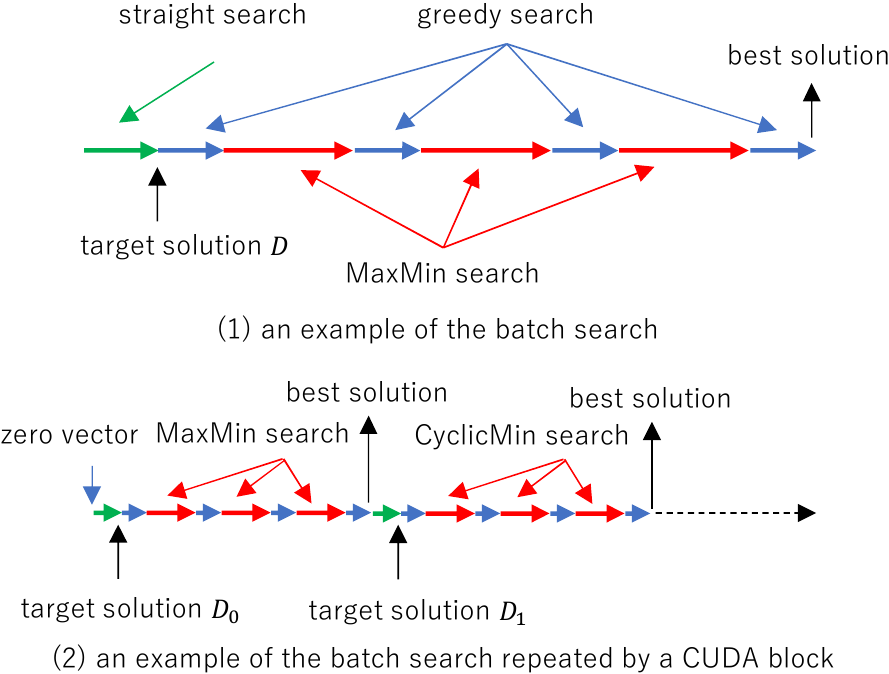}
\caption{Example of the batch search}
\label{fig:batch}
\end{figure}

Our DABS solver running on a GPU has multiple CUDA blocks (or thread blocks) of multiple threads.
Each CUDA block repeats \emph{the batch search}, which
maintains the current solution vector $X$ and performs one of the main search algorithms to traverse an $n$-bit search space
by repeatedly flipping a bit in $X$ by incremental search algorithms.
A target solution vector $D$ is given to a CUDA block, and
the Straight search destined for $D$ is performed.
Subsequently, it iterates the Greedy search algorithm and one of the main search algorithms as illustrated in Fig.~\ref{fig:batch}~(1).
In the figure, the MaxMin search is performed as one of the main search algorithms.
When the batch search terminates, the best solution obtained in Step~1 of the incremental search algorithm is output.
It makes no sense to repeat the TwoNeighbor search; hence, it is executed only once in a batch search.

In our GPU implementation of a QUBO solver, the host computes a target solution vector $D$, selects one of the main search algorithms,
and sends it to a GPU.
A CUDA block running on a GPU performs a batch search with the target solution vector and selected main search algorithm.
Fig.~\ref{fig:batch}~(2) illustrates how a CUDA block repeats the batch search. 
Initially, $X$ is a zero vector; hence, the energy $E(X)=0$.
In the figure, the CUDA block receives target solution vector $D_0$ and the MaxMin search algorithm is selected.
It performs a batch search with MaxMin and sends the best solution to the host PC when the batch search is finished.
Next, the CUDA block receives a packet with target solution vector $D_1$ and the CyclicMin search algorithm.
It performs the batch search in the same way.
The CUDA block repeats the batch search until it receives a termination request from the host.

We use two parameters, \emph{search flip factor} $s$ and \emph{batch flip factor} $b$ to determine number of flips (or iterations) performed
in the main search algorithm and batch search algorithm.
The main search algorithms, excluding the TwoNeighbor search algorithm, perform $sn$ flips, and the batch search algorithm terminates if the total number of flips exceeds $bn$.
For example,  if $n=1,000$, $s=0.6$, and $b=2.0$, then the main search algorithms performs $0.6\times 1,000=600$ flips and the batch search algorithm performs at least $2.0\times 1,000=2,000$ flips.
For simplicity of explanation, we assume that the Straight search algorithm performs 300 flips and each Greedy search performs 50 flips.
From $300+50+600+50+600+50+600+50=2,300\geq 2,000$,  the batch search performs the main search three times with 2300 total flips.

\subsection{Packets}
Communication between the host and GPUs in our DABS solver is done by packet transfer.
\emph{A packet} has four fields to store: a solution vector, the energy value, a main search algorithm, and a genetic operation,
as shown in Table~\ref{tab:packet}.
The fields of a packet sent from the host to a GPU include the following data. 
The solution vector field stores the target solution vector generated by the host.
The energy value field is void because the host never computes the energy.
The main search algorithm field is used to specify one of the main search algorithms to be executed by a CUDA block.
The genetic operation field is used to record one of the genetic operations performed to generate the target solution vector.

After a CUDA block receives a packet, it performs the batch search with the target solution vector and
the main search algorithm specified in the packet.
When the batch search terminates, a CUDA block overwrites the best solution in the solution vector field in the packet.
Moreover, the energy value of the best solution is written in the energy value field.
Subsequently, the packet is sent to the host.
Note that the fields for a main search algorithm and genetic operation are not updated.
The host uses these fields to record which main search algorithm and genetic operation were used to obtain the best solution.
The received packet is inserted into the solution pool if the best solution in the packet is better than the worst solution in the pool.

\begin{table}[!ht]
\centering
\caption{Examples of packets}
\label{tab:packet}
\includegraphics{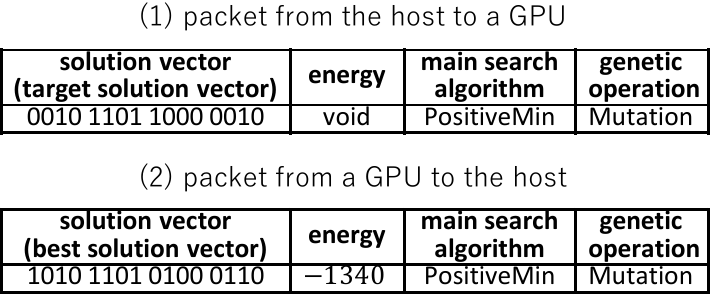}
\end{table}

\section{GA-based search algorithm}
\label{sec:GA}

This section explains the GA-based search algorithm for solving QUBO models.
The batch search is executed by CUDA blocks on GPUs, whereas the GA-based search algorithm runs on the host.
The GA-based search algorithm has \emph{a solution pool}, which stores packets with good solution vectors obtained by the batch search.
Fig.~\ref{fig:outline} shows an example of a solution pool of size 5, which can record 5 packets received from a CUDA block.


\subsection{Genetic operations}
We use several genetic operations to generate diverse target solutions.
For this purpose, we may pick at most two solutions from a solution pool at random,
and genetic operations are applied to them to generate a target solution.
It makes sense to select a better solution in the solution pool with a higher probability.
We generate a uniform random number $r$ in $[0,1)$ and select the $(\lfloor r^3\cdot m\rfloor +1)$-th
solution vector in the solution pool, where $m$ is the number of solution vectors in it.
For example, the first solution is selected with probability $1\over\sqrt[3]{m}$, which is higher than equal probability $1\over m$.

We use the following genetic operations for selected solutions.\\
\emph{Mutation}:
We use one randomly selected solution vector, and each bit of it is flipped with some small fixed probability $p$, say $1\over 8$.\\
\emph{Crossover}:
Two randomly selected vectors are mixed such that each bit of a target solution vector is selected from the corresponding bit of one of the two selected vectors at random.\\
\emph{Zero}:
We use one randomly selected solution vector, and 0 is written to each bit with some fixed probability $p$, say $1\over 8$.\\
\emph{One}:
Similarly to Zero above, 1 is written to each bit with some small fixed probability.\\
\emph{IntervalZero}:
We use one randomly selected solution vector.
A segment of the vector is chosen at random, and 0 is written to all bits in it.
For example, the size of a segment is determined by a random integer in $[32,n/2]$,
and the segment is arranged to the $n$-bit solution vector.
All bits in the segment are set to 0.\\
\emph{Best}:
The best solution in the solution pool is used as-is.\\
\emph{Random}:
We simply generate a random solution vector without using a solution in the solution pool.\\

The host selects one of the genetic operations and one of the main search algorithms as follows.
Initially, a solution pool in the host memory is filled with random solution vectors with $+\infty$ energy.
Moreover, columns of the main search algorithm and the genetic operations are initialized at random.
The host repeatedly generates target solutions by genetic operations.
With small probability, say, 5\%, one of the genetic operations is selected uniformly at random.
With high probability, say 95\%, one of the rows of the solution pool is selected uniformly at random,
and the genetic operation written in the selected row is used.
Because genetic operations by which good solution vectors have been obtained are written in the solution pool,
such genetic operations are selected with a higher probability.
Similarly, one of the main search algorithms is selected.
With a small probability, say, 5\%, one of the main search algorithms is selected uniformly at random.
With high probability, say 95\%, one of the main search algorithms in the solution pool is selected
uniformly at random.
The selected main search algorithm and genetic operation are written in a new packet.
Moreover, a new solution vector generated by the selected genetic operation is written in the new packet.
The host sends this new packet to the associated GPU and a CUDA block performs the batch search base on it.

\subsection{Island model}
A solution pool tends to be filled with relatives of the best solution vector.
Low diversity of solution vectors in the solution pool degrades the search performance,
because the batch search traverses the $n$-bit search space close to the best solution vector.
We can use the island model~\cite{Whitley98} for a higher diversity of the GA.
In the island model, multiple solution pools analogous to islands are used.
Fig.~\ref{fig:outline} illustrates the island model with four solution pools.
We assume that solution pools have the cyclic order shown in the figure.

In conventional island models, solution migration between solution pools is performed.
However, our DABS solver does not perform solution migration.
Instead, a genetic operation called {\em Xrossover} (or inter-pool crossover) is performed as follows.
Two solution vectors are picked, one from the associated solution pool and the other from a neighbor solution pool.
The Crossover genetic operation is performed for these two solutions to obtain a target solution.
We can consider this target solution as a midway point between these two solution pools.
The batch search is performed for this midway target solution.
The batch search may find a very good solution near the midway target solution,
which will be inserted in the solution pool.
Hence, the batch search for a target solution vector obtained by the Xrossover traverses from the source solution pool to the midway solution,
and the search space can be expanded.
Further,  because the best solution obtained from the midway target solution is inserted into the destination solution pool,
we can consider the source solution pool moves slightly toward the destination solution pool in the $n$-bit search space.
Thus, if the Xrossover operation succeeds in finding good solutions many times, then the ``ring'' of solution pools  in Fig.~\ref{fig:outline} is reduced,
and all solution pools may be merged into a very good solution, in the sense that all solutions in all solution pools are similar.
However, after all solution pools are merged, we have very few chances to find a better solution because all solutions are relatives.
If this is the case, we can initialize all solution pools as random solution vectors and restart the QUBO solver from the beginning to find a better solution.

\section{GPU implementation of our DABS solver}
\label{sec:GPU}
This section shows how our QUBO solver is implemented to run on a host with multiple GPUs using
CUDA C++~\cite{CUDA-Programming} with Open MP.
We assume that we use eight NVIDIA A100 GPUs (Ampere Architecture, Compute Capability 8.0)~\cite{NVIDIA-A100}
equipped with a global memory of size 40~GB.
It has 108 Streaming Multiprocessors (SMs), each of which has a 256~KB register file, and 192~KB of combined shared memory and L1 data cache.
The matrix $W=(W_{i,j})$ of a QUBO model is stored in the global memory of GPUs.
CUDA blocks running on GPUs perform the batch search.
We store the current vector $X=(x_k)$ and $\Delta_k(X)$ in the register of threads in a CUDA block.
Since a CUDA block can have up to 1024 threads, we use a CUDA bock with $n$ threads only if $n\leq 1024$,
and each thread $k$ ($0\leq k\leq n-1$) stores the value of bit $x_k$ and $\Delta_k(X)$ in its registers.
If $n>1024$ then each thread stores the values of $n\over 1024$ bits and $n\over 1024$ $\Delta_k(X)$'s.
Because each SM can have up to 2048 resident threads,
it can load two CUDA blocks with 1024 threads; thus, $108\cdot 2=216$ CUDA blocks can be dispatched in an NVIDIA A100 GPU.

We will show how the batch search is implemented to run in a GPU.
The values of ${\it BEST}$ and $E({\it BEST})$ are stored in the shared memory of a GPU,
which is a low-latency memory that can be accessed by threads in the same CUDA block.
We use a CUDA atomic function atomicMin~\cite{CUDA-Programming} to find the minimum value.
The readers may think that reduction operations should be used to find the minimum value efficiently.
However, atomicMin can be much more efficient than the conventional reduction operation to implement Step~1 of the incremental search algorithm.
Each thread computes the value of $E(f_j(X))$, and if $E({\it BEST})>E(f_j(X))$, then ${\rm atomicMin}(\&x,E(f_j(X)))$ is performed.
We assume that $x$ is initialized by $+\infty$.
If $x=+\infty$ is satisfied after this operation, then no thread has executed atomicMin and no 1-bit neighbor solution is better than ${\it BEST}$.
Otherwise, $E({\it BEST})>x$ holds, and we update ${\it BEST}$ by $f_j(X)$ satisfying $x=E(f_j(X))$.
Because ${\rm atomicMin}(\&x,E(f_j(X)))$ is performed only if $E({\it BEST})>E(f_j(X))$, that is,
only if we can obtain a new best solution by flipping a bit,
the frequency of updating the best solution is relatively small.
Hence, with high probability, no thread performs ${\rm atomicMin}(x,E(f_j(X)))$,
and thus updating ${\it BEST}$ by atomicMin can be much more efficient than the conventional reduction operation.

Eight Open MP threads are used to invoke CUDA kernels for the batch search on eight GPUs.
Moreover, additional eight Open MP threads are used to operate on eight solution pools.
This maintains the assigned solution pool, and generates packets to be sent to the GPU.

Some of the main search algorithms require random numbers.
For random number generation on the GPU, the host generates random seeds using the Mersenne twister~\cite{Matsumoto98}
and transfers them to all threads of the GPUs through the global memory such that each thread has a 64-bit random seed.
Each thread performs Xorshift~\cite{Marsaglia03} to generate new random numbers from the 64-bit random seed quickly.


\section{Experimental results}
\label{sec:experiment}
We used commercially available solvers against which we compared the performance of our DABS solver.
First, we used Gurobi 9.5.1~\cite{Gurobi}, an MIP solver supporting quadratic formulas.
For solving QUBO models by Gurobi, a computing server equipped with two AMD EPYC 7702 64-core processors (2GHz) and 2~Tbyte memory
was used.
We set ${\rm MIPFocus}=1$ to focus on finding feasible solutions, ${\rm Threads}=256$ to exploit up to 256 logical cores of 128 physical cores,
and ${\rm TimeLimit}=3,600$s.

We also used a quantum annealer D-Wave Advantage4.1~\cite{Advantage} for solving the QASP.
Further, a D-Wave Hybrid solver~\cite{D-Wave-Hybrid20} running on a D-Wave Cloud service was used
for solving QUBO problems reduced from the MaxCut problems and QAPs.

For evaluating the performance of our DABS solver, we used a compute node of AI Bridging Cloud Infrastructure (ABCI) supercomputer~\cite{ABCI2.0}
equipped with two Intel Xeon Platinum 8360Y CPUs (2.40~GHz), 512~GB memory, and eight NVIDIA A100 GPUs.
Our DABS solver is developed by CUDA C++ in CUDA compilation tools, release 11.6.
We use solution pools with 100-packet capacity, and the tabu period is fixed to 8.

For fair evaluation of the ABS solver~\cite{Yasudo-JPDC22}, we used the same compute node.
Owing to the low diverse, the ABS solver may stack at a non-optimal local minimum and no better solution cannot be found.
Hence, even if the ABS can find a potential optimal solution in few seconds with high probability, it cannot find it in several hours with small probability.
It is not appropriate to evaluate it only by the TTS to obtain potential optimal solutions.
Thus, we set a time limit for the ABS, and evaluate the probability to obtain a potential optimal solution within the time limit
and the TTS to obtain the potential optimal solution.
Note that the TTS does not count the execution time of a trial if it fails to find the potential optimal solution within the time limit.

Throughout this section, several histograms are used to illustrate the experimental results.
In these histograms, bins with labels $b_1, b_2, \ldots$ mean that each $b_i$ ($i=1, 2,\ldots $) corresponds to the range $[b_i,b_{i+1})$.

\subsection{MaxCut problem}
We used three 2000-node graphs, K2000, G22, and G39, for benchmarking.
K2000 is a randomly generated complete graph with weights $\{-1,+1\}$~\cite{k2000}.
G22 and G39 are sparse graphs in the MaxCut problem collection Gset~\cite{gset},
with weights $\{+1\}$ and $\{-1,+1\}$, respectively.
These graphs are also used to evaluate an Ising model solver called \emph{Coherent Ising Machine (CIM)}~\cite{Inagaki16},
which takes an optical processing approach based on a network of coupled optical pulses in a 1-km ring fiber measured and controlled by an FPGA module.

\begin{figure}[!ht]
\centering
\includegraphics[width=8cm]{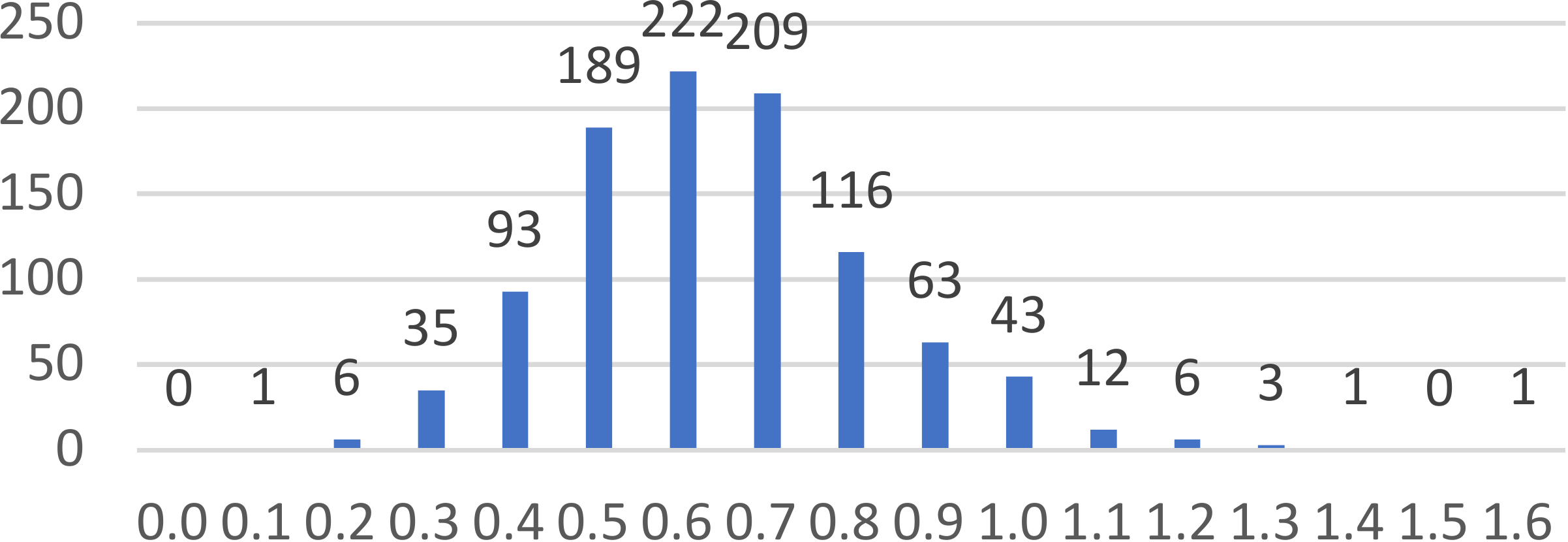}
\caption{Histogram of Time-To-Solution (TTS) in seconds for 1,000 executions of our DABS solver to obtain the potentially optimal solution $-3,3337$}
\label{fig:maxcut_gpu}
\end{figure}

Table~\ref{tab:maxcut} summarizes the experimental results.
We explain this table using MaxCut K2000 as an example because experimental results for the three graphs have the same tendency.
We executed our DABS solver 1,000 times with search flip factor $s=0.1$ and batch flip factor $b=10$ for K2000.
It succeeded in finding the potentially optimal solution of $-3,3337$ for all 1,000 executions.
The average TTS for the potentially optimal solution is only 0.694s.
Fig.~\ref{fig:maxcut_gpu} shows the histogram of the TTS to obtain the potentially optimal solution.
We can see that the TTS of all 1,000 executions is less than 1.7s.

\begin{table}
\centering
\caption{Experimental results for the MaxCut problems, K2000, G22, and G39}
\label{tab:maxcut}
\begin{tabular}{lrrr}
   MaxCut          & K2000 & G22 & G39 \\ 

Potential optimal solution &  $-33,337$ & $-13,359$ & $-2,408$ \\
\hline
Our DABS solver  & $-33,337$  & $-13,359$ & $-2,408$\\
\hfill (TTS) & 0.694s  & 1.58s &   7.56s \\
\hline
ABS solver & $-33,337$  & $-13,359$ & $-2,408$\\
\hfill (TTS) & 9.19s  & 19.7s &   15.1s \\
\hfill (Probability) & 99.2\%  & 69.5\% &  78.6\% \\
 \hline
Gurobi optimizer & $-33,241$ & $-13,137$ & $-2,276$\\
 \hfill (Gap) & 0.287\% &  1.66\% & 5.48\%\\
\hline
D-Wave Hybrid solver& $-33,337$ & $-13,359$ & $-2,408$  \\
 \hfill       (TTS) & 100-200s & 10-20s & 50-100s\\
 \hline 
 CIM~\cite{Inagaki16}   & $-33,191$ & $-13,313$ & $-2,361$ \\
 \hfill (Gap) & 0.438\% & 0.344\% & 1.95\%
\end{tabular}
\end{table}

Gurobi optimizer was executed for 3,600s, but it failed to find the potentially optimal solution.
It found a solution $-33,241$, which is 0.287\% gap from the potentially optimal solution $-33,337$.

\begin{figure}[!ht]
\centering
\includegraphics[width=8cm]{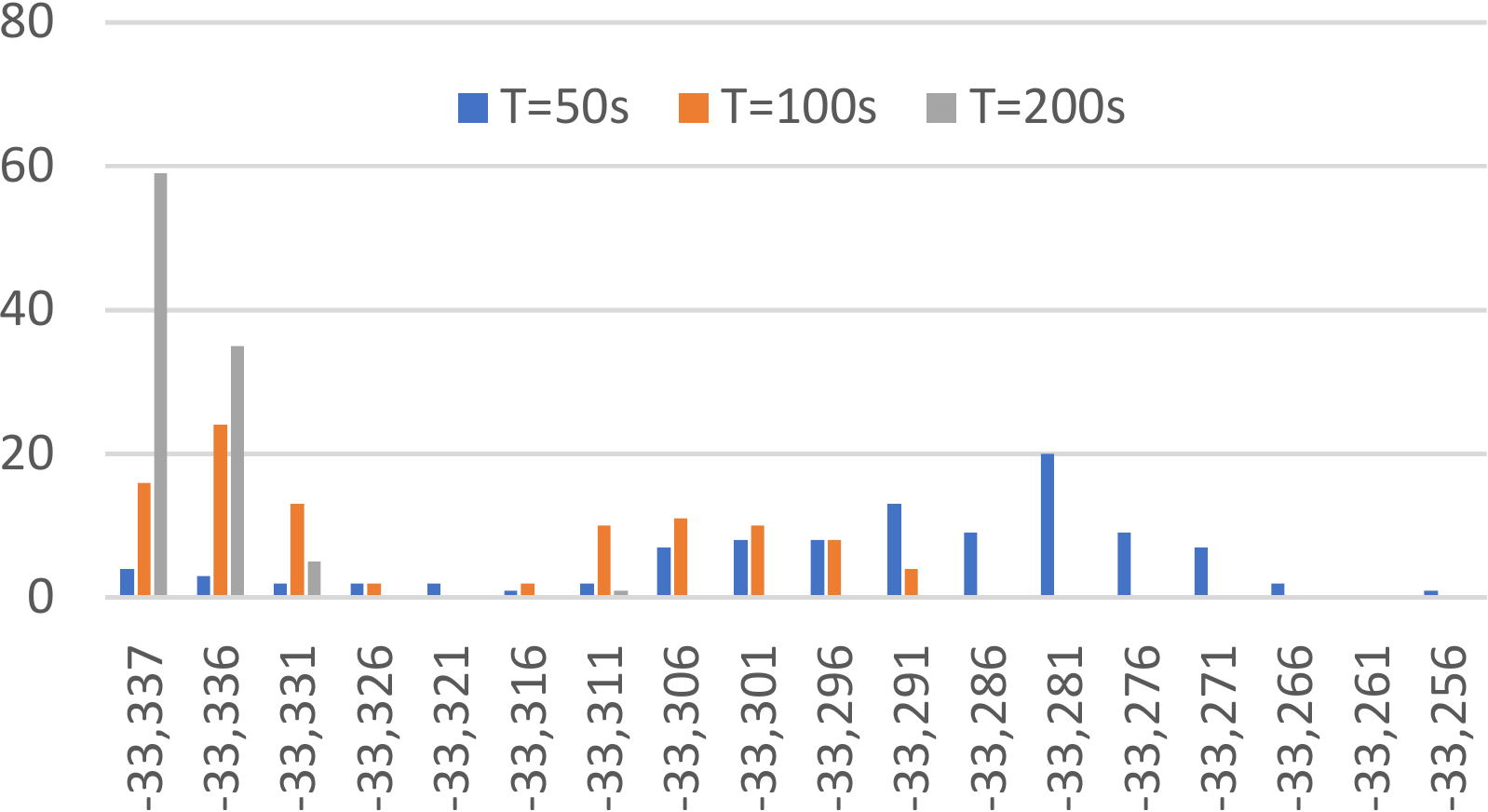}
\caption{Histogram of solutions for K2000 MaxCut obtained in 100 executions of D-Wave Hybrid Solver running $T=50$s, $T=100$s, and $T=200$s. }
\label{fig:k2000_dwave}
\end{figure}

Meanwhile, D-Wave Hybrid Solver succeeded in finding the potentially optimal solution.
However, the D-Wave API library does not have a function to evaluate the TTS to obtain a particular solution,
and it just outputs the best solution within a fixed time limit.
Hence, we executed it for fixed time limits $T=50$s, $T=100$s, and $T=200$s
to estimate the TTS.
Fig.~\ref{fig:k2000_dwave} shows the histogram of solutions in 100 executions of the D-Wave Hybrid solver.
It found a potentially optimal solution $-33,337$ for 4, 16, and 59 times out of 100 executions each for $T=50$s, $T=100$s, and $T=200$s, respectively.
Hence, we can say that the TTS to obtain the potentially optimal solution was between 100s and 200s.


The K2000 graph is commonly used for benchmarking BQM solvers for solving MaxCut problems.
It has been reported that the best solution obtained in 100 executions of 5~ms each of the CIM is $-33,191$~\cite{Inagaki16}.
The gap between this best solution and the potential optimal solution is 0.438\%.
A 512-spin annealing processor fabricated with the 65~nm CMOS technology has been developed~\cite{Yamamoto21}.
They estimated that numerical projections for a 2048-spin annealing processor obtain expected -33,073 solutions in 100 trials of 0.48~ms each for the K2000 problem.
However, it fails to find a solution better than -33,300.
A new optimization algorithm called \emph{the Simulated Bifurcation Machine (SBM)},
which simulates adiabatic evolution of classical nonlinear Hamiltonian systems for solving Ising models,
has been developed~\cite{Goto19}.
It has been implemented in an FPGA to solve 2000-node Ising models
and 100 trials of 0.5~ms running time for the K2000 problem obtained solutions with an average $-32,768$.
However, it failed to obtain solutions better than $-33,000$.
Recently, the SBM ~\cite{Goto19} has been improved;
An FPGA implementation of the improved version of the SBM called \emph {discrete Simulated Bifurcation (dSB)}
succeeded in finding a potentially optimal solution $-33,337$ in 1.3s~\cite{Goto21}.
Our DABS solver running on GPUs can find it faster than this dSB FPGA solver.


\subsection{Quadratic Assignment Problem (QAP)}
We used three QAP problems, tai20a ($n=20$), tho30 ($n=30$), and nug30 ($n=30$), from QAPLIB, which is a quadratic assignment problem library~\cite{Burkard91}.
Their optimal solutions have been proved.
QAP problems tai20a and tho30 have been used for evaluating a QUBO solver using the subQUBO model~\cite{Atobe21}.
However, this solver failed to find optimal solutions.
We also used nug30, which is the largest QAP of commonly used nug-family QAPs.
Table~\ref{tab:exQAP} lists the experimental results for QUBO problems reduced from tai20a, tho30, and nug30.
We explain this table using nug30 as an example. 
It is known that the optimal solution of QAP nug30 is $6,124$.
We used penalty $p=1,000$, and so the optimal solution of the corresponding QUBO problem is
$6,124-np=-23,876$.
Our DABS solver with parameters $s=0.1$ and $b=1$ succeeded in finding this optimal solution for all 1,000 executions,
and the average TTS was 44.2s.
Gurobi optimizer found a solution $-23,820$ in 3,600s, which is 0.235\% gap from the optimal solution.
Our DABS solver could find the optimal solution in all 1,000 executions;
however, the ABS solver can find it with probability only 14.8\% in a time limit of 300s,
and Gurobi optimizer and D-Wave Hybrid solver failed to find it.

\begin{table}
\caption{Experimental results for QAPs, tai20a, tho30, and nug30}
\label{tab:exQAP}
\begin{tabular}{lrrr}
           QAP           & tai20a & tho30 & nug30\\ 

QAP optimal solution  & $703,482$ & $149,936$ & $6,124$\\
Penalty &  $200,000$ & $30,000$ & $1,000$ \\
QUBO optimal solution  & $-3,296,518	$&$-750,064$ &$-23,876$\\
 \hline
Our DABS solver  & $-3,296,518	$&$-750,064$ &$-23,876$\\
\hfill (TTS) & 81.6s & 9.60s & 44.2s \\
\hline
ABS solver  & $-3,296,518	$&$-750,064$ &$-23,872$\\
\hfill (TTS) & 93.5s & 38.6s& 51.7s\\
\hfill (Probability) & 13.4\% & 67.5\%&14.8\% \\
\hline
Gurobi optimizer  & $-3,291,532$ & $-749,034$ & $-23,820$ \\
 \hfill (Gap) & 0.151\% & 0.137\% &0.235\%\\
\hline
D-Wave Hybrid solver&$-3,235,364$ & $-738,136$	 & $-23,350$ \\
 \hfill (Gap) & 1.86\% &  1.59\% &  2.20\% \\
\end{tabular}
\end{table}

\subsection{Quantum Annealer Simulation Problem (QASP)}
We used QASPs for D-Wave Advantage 4.1 with resolutions $r=1$, $r=16$, and $r=256$ for benchmarking.
Ising models for the QASPs have 5,627 nodes and 40,279 edges.
The values of interactions $J_{i,j}$ and biases $h_{i}$ are randomly selected from all possible non-zero values.
For example, for generating a QASP with $r=1$, each $J_{i,j}$ is selected from $\{-1,+1\}$ and each $h_{i}$ is selected
from $\{-4, -3, -2, -1, +1, +2, +3, +4\}$, with equal probability.
Ising models with $J_{i,j}$'s and $h_{i,j}$'s obtained are thus converted to the equivalent QUBO models
for our DABS solver and Gurobi optimizer. 
We write QASP1, QASP16, and QASP256 to denote these QASPs with resolutions $r=1$, $r=16$, and $r=256$, respectively.

\begin{table}
\caption{Experimental results for QASP1, QASP16, and QASP256, }
\label{tab:qasp}
\begin{tabular}{lrrr}
      QASP         & QASP1 & QASP16 & QASP256 \\ 
      resolution   & $r=1$ & $r=16$ & $r=256$ \\
Potentially optimal solution & $-20,902$ & $-238,594$ & $-3,656,992$\\
 \hline
Our DABS solver  & $-20,902$ & $-238,594$ & $-3,656,992$\\
\hfill (TTS) &4.34s & 5.67s & 5.33s \\
\hline
ABS solver  & $-20,902$ & $-238,594$ & $-3,656,982$\\
\hfill      (TTS)         & 6.92s & 12.16s & 4.57s\\
\hfill (Probability) & 93.2\% & 18.6\%&28.3\%\\
\hline
Gurobi optimizer& $-20,676$ & $-238,582$ & $-3,656,192$\\
 \hfill (Gap) & 1.08\% & 0.00503\% &0.0219\%\\
\hline
D-Wave Advantage \ & $-20,880$& $-238,430$ & $-3,654,338$ \\
 \hfill (Gap) & 0.105\% &  0.0687\% &  0.0726\% \\
\end{tabular}
\end{table}

\begin{figure}[!ht]
\centering
\includegraphics[width=8cm]{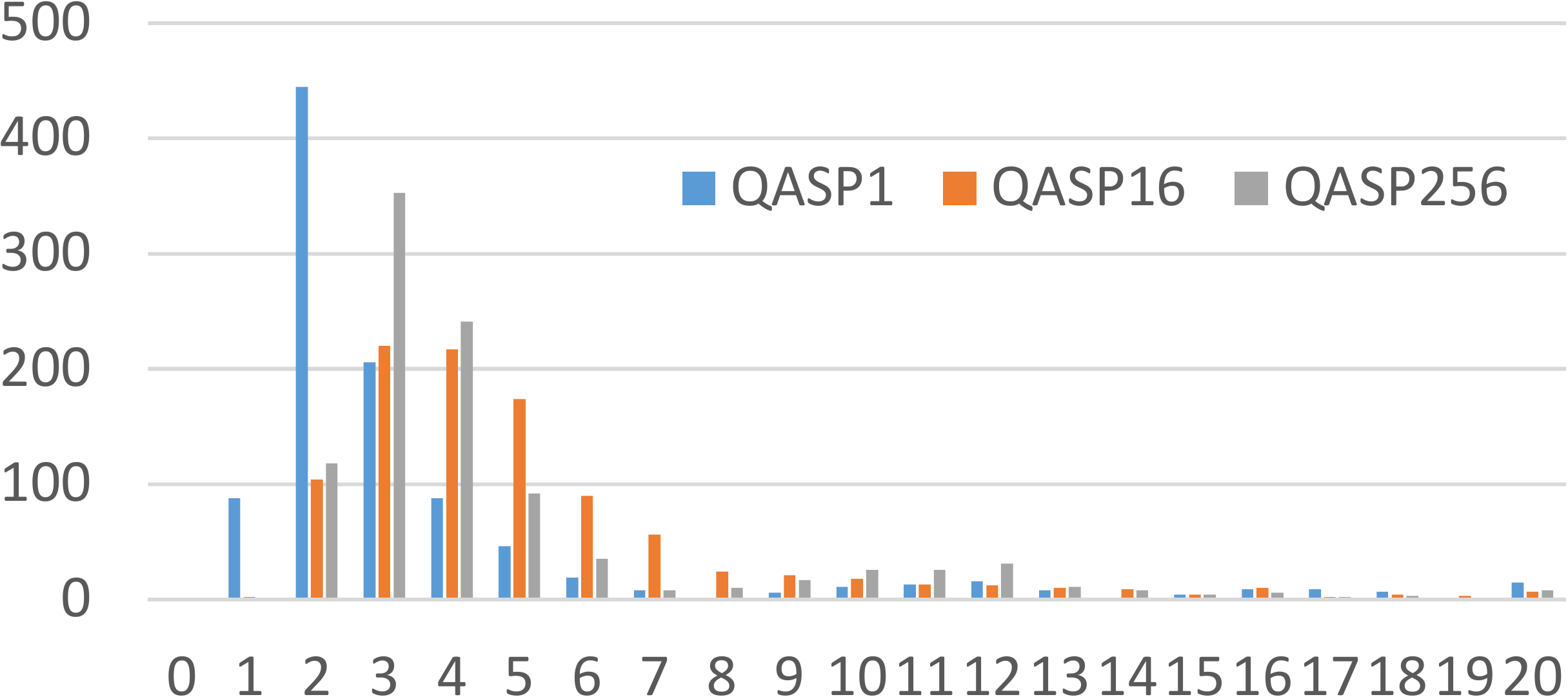}
\caption{Histogram of running time in seconds for 1,000 executions of our DABS solver to obtain the potentially optimal solutions $-20,902$, $-238,594$, and $-3,656,992$
of QASP1, QASP16, and QASP256, respectively}
\label{fig:qasp_gpu}
\end{figure}

Table~\ref{tab:qasp} summarises the experimental results for the QASPs.
Our DABS solver with parameters $s=0.1$ and $b=1$ obtained potentially optimal solutions in all $1,000$ executions for each resolution,
and the average TTSs were 4.34-5.67s.
Fig.~\ref{fig:qasp_gpu} shows the histogram of the running time to obtain the potentially optimal solutions.
We can see that the optimal solutions were obtained in less than 10s with high probability for all QASPs.
We executed ABS solver with a time limit of 30s for solving QASPs.
For example, it succeeded in finding the potentially optimal solutions with probabilities of 93.2\%, 18.6\%, and 28.3\% in this time limit, respectively.
For QASP256, the TTS of the ABS solver is smaller than that of our DABS solver.
Hence, the reader may think the ABS is better than the DABS solver for this case.
However, the ABS solver succeeded in finding the potentially optimal solution with probability only 28.3\%,
while the probability of our DABS solver is 100\%.

Gurobi optimizer was executed for 3,600s for solving the QASPs.
The obtained solutions were close to the potentially optimal solutions; however, there were few gaps.

Table~\ref{tab:qasp} also shows the best solution obtained by 1,000,000 quantum annealing operations with 20~$\mu$s annealing time each by the D-Wave Advantage4.1.
Because one API call called \emph{sampling} can perform at most 10,000 quantum annealing operations, we repeated the sampling 100 times.
Including miscellaneous overhead, each sampling with 10,000 quantum annealing takes about 2.7s, and the total execution time for 100-time sampling is about 37min.
For comparison, the energy values in the table are converted from the Hamiltonian values of the corresponding Ising models solved by D-Wave Advantage4.1.
For all QASPs, it failed to find the potentially optimal solutions in all 1,000,000 quantum annealing operations.



\subsection{Effect of Diversity for Main Search Algorithms and Genetic Operations}
Recall that our DABS solver performs five main algorithms, MaxMin, PositiveMin, CyclicMin, RandomMin, and TwoNeighbor,
and eight genetic operations, Random, Best, Mutation, Crossover, Xrossover, Zero, One, and IntervalZero.
The host chooses a main algorithm and genetic operation based on solutions stored in the solution pool
so that a main algorithm and a genetic operation that have produced good solutions are selected more frequently.
To see the effect of the diversity of main search algorithms and genetic operations,
we evaluated the frequency of main search algorithms and genetic operations used during the execution of our DABS solver.
We executed our DABS solver 1,000 times each for solving QUBO problems to obtain the experimental results in Tables~\ref{tab:maxcut}, \ref{tab:exQAP}, and~\ref{tab:qasp}.
During the execution, it recorded the frequency of executed main search algorithms and genetic operations.
Table~\ref{tab:freq1} shows the frequency of main search algorithms and genetic operations executed in our DABS solver for each QUBO model.
For each QUBO model in the table, the main search algorithm and genetic operation with the highest frequency are boldfaced.
For example, for solving a QUBO problem reduced from nug30,
PositiveMin search algorithm and Crossover genetic operation are selected most frequently, and their frequencies
are 44.9\% and 62.8\%, respectively.
We can consider that main search algorithms/genetic operations with higher frequency found better solutions for a QUBO model in the experiments.
Even if QUBO models are reduced from the same problem category,  the most frequently used main algorithm/genetic operation can be different.
For example, PositiveMin is used most frequently for solving QASP1 and QASP16, whereas CyclicMin is mostly selected for QASP256.
Although their difference is just the resolution, the main search algorithm that works well is not the same.
We are not able to clarify the reason for this, but our diverse approach of automatic selection of main algorithms/genetic operations may work well.

\begin{table*}
\centering
\caption{Frequency of main search algorithms and genetic operations executed in our DABS solver}
\label{tab:freq1}
\begin{tabular}{c|ccccc|cccccccc}
  &\multicolumn{5}{c|}{Main search algorithms} &\multicolumn{8}{c}{Genetic operations}\\
Problems  &Max &Positive& Cyclic & Random &Two  & Ran- & Best & Muta- & Cross- & Xross- & Zero & One & Interval\\ 
 & Min  &  Min & Min & Min &Neighbor & dom & & tion & over & over & & & Zero\\   
\hline
K2000 & 15.7\% & \bf{25.1\%} & 20.7\% & 24.8\% & 13.7\% &\bf{26.4\%} & 10.7\% & 11.0\% & 9.4\% & 8.6\% & 11.0\% & 11.0\% &11.9\%\\
G22 & 10.4\% & \bf{31.8\%} & 15.2\% & 29.3\% & 13.3\% &  14.9\% &9.9\% &14.1\% &10.8\% &7.4\% &13.4\% &13.5\% &\bf{16.0\%}\\
G39 & 5.5\% & 18.3\% &	15.9\% & \bf{44.6\%} &	15.7\% &6.5\%& \bf{17.9}\% &	16.3\% & 14.3\% &	2.5\% &16.1\% &16.2\% &	10.2\% \\
\hline
tai20a &12.7\% &\bf{60.4\%} &10.0\% &10.1\% &6.8\% & 0.4\% & 1.7\% &3.6\% &5.0\% &0.9\% &\bf{73.0\%} &2.0\% &13.4\%\\
tho30 &14.9\% &\bf{58.4\%} &10.9\% & 11.4\%& 4.4\% & 0.5\% &1.7\% &1.2\% & 39.6\% &1.0\% & \bf{49.3\%} &	1.1\% & 5.6\%\\
nug30 & 20.0\%  &\bf{44.9\%}	&14.4\% &15.1\% & 5.6\% & 0.2\% & 1.3\%& 1.1\% &	\bf{62.8\%} & 1.0\% &27.8\% &	 1.0\% & 4.8\%\\
\hline
QASP1 & 4.8\%  &\bf{46.1\%}  &12.9\%  &33.1\%  &3.1\%  & 3.8\% &16.1\% &9.2\% &\bf{41.1\%} &7.4\% &10.2\% &9.5\% &2.7\%\\
QASP16& 5.8\%  & \bf{37.5\%}  &20.1\%  &27.9\%  &8.7\%  &2.4\%&5.4\%&11.9\%&26.6\%&\bf{28.0\%}&12.8\%&10.8\%&2.1\%\\
QASP256 &8.5\%  &16.2\%  &\bf{35.7\%}  &30.5\%  &9.1\%  &2.5\%&6.0\%&11.5\%&27.7\%&\bf{31.3\%}&9.2\%&9.5\%&2.3\%\\
\end{tabular}
\end{table*}

Our DABS solver records a main search algorithm and genetic operation used for a batch search when it finds a new best solution.
This record is updated whenever a new best solution is found.
Hence, when our DABS solver terminates, we can obtain the main search algorithm/genetic operation by which the potentially optimal solution was found first.
Table~\ref{tab:freq2} lists the frequency of main search algorithms/genetic operations obtained by reading this record.
Intuitively, a higher frequency of a main search algorithm/genetic operation means that it works well for finding a potentially optimal solution
from good solutions.
Notably, the frequencies shown in Tables~\ref{tab:freq1} and~\ref{tab:freq2} do not show the exact same tendency.
For example, for QASP256, CyclicMin is mostly used in Table~\ref{tab:freq1}, whereas RandomMin is mostly selected in Table~\ref{tab:freq2}.
This means that for solving QASP256, CyclicMin works well for finding good solutions,
while RandomMin is most suitable for finding the potentially optimal solution from good solutions.
This fact implies that the best main search algorithm may be changed during the execution of a QUBO solver,
and the performance can be improved if we can select a main search algorithm appropriately.
Moreover, for MaxCut problems, K2000, G22, and G39, genetic operation Best works well to find potentially optimal solutions, as shown  
in Table~\ref{tab:freq2}.
However, Best is not mostly used for finding good solutions as we can see in Table~\ref{tab:freq1}.
Our DABS solver is designed so that suitable main search algorithms/genetic operations are automatically selected and executed.

\begin{table*}
\centering
\caption{Frequency of main search algorithms and genetic operations that firstly find the potentially optimal solutions}
\label{tab:freq2}
\begin{tabular}{c|ccccc|cccccccc}
&\multicolumn{5}{c|}{Main search algorithms} &\multicolumn{8}{c}{Genetic operations}\\
Problems  &Max &Positive& Cyclic & Random &Two  & Ran- & Best & Muta- & Cross- & Xross- & Zero & One & Interval\\ 
 & Min  &  Min & Min & Min &Neighbor & dom & & tion & over & over & & & Zero\\    
\hline
K2000 & 0.8\% &\bf{93.1\%} &0.4\% &5.6\% &0.1\%& 0.1\%&	\bf{36.5\%}&14.9\%&5.5\%&1.3	\%&14	.0\%&10.7\%&17.0\% \\
G22 & 2.5\% &\bf{69.9\%} &3.6\% &16.7\% &7.3\% &0.0\%& 21.7\%&16.9\%&5.4\%&0.5\%&14.6\%&18.4\%&\bf{22.5\%}\\
G39 & 0.6\% &27.2\% &	2.1\% &\bf{66.0\%} &	4.1\% &0.0\% & \bf{42.6\%} & 13.5\% & 	8.4\% & 0.4\% & 13.5\% & 14.8\% & 6.8\% \\
\hline
tai20a & 9.6\% &\bf{78.4\%} &4.5\% &5.2\% &2.3\% &0.0\%&2.3\%&3.5	\%&5.0\%&0.1\%&\bf{75.5\%}&0.3\%&13.3\%\\
tho30 & 15.9	\% &\bf{60.3\%} &10.2\% &9.8\% &3.8\% & 0.0\%&2.9\%&0.2\%&\bf{60.5\%}&0.1\%&34.9\%&0.0\%&1.4\%\\
nug30 &21.2	\% &\bf{51.1\%} &11.8\% &11.2\% &4.7\% &0.0\%&0.4\%&0.0\%&\bf{75.5\%}&0.1\%&23.3\%&0.1\%&0.6\%\\
\hline
QASP1 & 0.9\% &\bf{49.9\%} &1.5\% &46.3\% &1.4\% &0.0\%&16.5\%&5.9\%& \bf{60.2\%}&2.9\%&6.4\%&7.9\%&0.2\%\\
QASP16& 3.7\% &\bf{43.7\%} &5.0\% &12.8\% &34.8\% & 0.0\%&16.0\%&10.8\%&\bf{30.3\%}&20.1\%&3.6\%&19.0\%&0.2\%\\
QASP256&10.1\% &18.2\% &17.5\% &\bf{30.9\%} &	23.3\% &0.0\%&1.8\%&19.6\%&24.6\%&	\bf{38.8\%}&	14.4\%&0.8\%&0.0\%
\end{tabular}
\end{table*}

\section{Conclusion}
\label{sec:concl}
This paper presented a QUBO solver framework called Diverse Adaptive Bulk Search (DABS), which features three diverse aspects: five search algorithms, eight multiple genetic operations, and multiple solution pools.
During the execution of the DABS, search algorithms and genetic operations that succeeded in finding good solutions
are automatically selected to obtain better solutions more quickly.
Although the well-known No Free Lunch Theorem (NFLT) implies that there exists no heuristic search algorithm
that can solve all types of QUBO problems efficiently, our DABS solver has the potential to solve many types of QUBO problems without knowing their characteristics.
We implemented a QUBO solver based on the DABS to run on the compute node of the ABCI supercomputer equipped with eight NVIDIA A100 GPUs.
Experimental evaluations confirm that our DABS solver succeeds in finding optimal or potentially optimal solutions for
three types of QUBO problems, the MaxCut problem, the Quadratic Assignment Problem (QAP),
and Quantum Annealer Simulation Problem (QASP).
\bibliographystyle{IEEEtran}
\bibliography{algorithm,gpu,fpga,sort,mining}

\begin{thebibliography}{10}
\providecommand{\url}[1]{#1}
\csname url@samestyle\endcsname
\providecommand{\newblock}{\relax}
\providecommand{\bibinfo}[2]{#2}
\providecommand{\BIBentrySTDinterwordspacing}{\spaceskip=0pt\relax}
\providecommand{\BIBentryALTinterwordstretchfactor}{4}
\providecommand{\BIBentryALTinterwordspacing}{\spaceskip=\fontdimen2\font plus
\BIBentryALTinterwordstretchfactor\fontdimen3\font minus
  \fontdimen4\font\relax}
\providecommand{\BIBforeignlanguage}[2]{{%
\expandafter\ifx\csname l@#1\endcsname\relax
\typeout{** WARNING: IEEEtran.bst: No hyphenation pattern has been}%
\typeout{** loaded for the language `#1'. Using the pattern for}%
\typeout{** the default language instead.}%
\else
\language=\csname l@#1\endcsname
\fi
#2}}
\providecommand{\BIBdecl}{\relax}
\BIBdecl

\bibitem{Kadowaki98}
T.~Kadowaki and H.~Nishimori, ``Quantum annealing in the transverse {Ising}
  model,'' \emph{PHYSICAL REVIEW E}, vol. E58, no.~5, pp. 5355--5363, Nov.
  1998.

\bibitem{Brush67}
S.~G. Brush, ``History of the {Lenz-Ising} model,'' \emph{Rev. Mod. Phys.}, p.
  883, Oct. 1967.

\bibitem{McGeoch19}
C.~C. McGeoch, R.~Harris, S.~P. Reinhardt, and P.~Bunyk, ``Practical
  annealing-based quantum computing,'' \emph{IEEE Computer}, vol.~52, pp.
  38--46, Jun. 2019.

\bibitem{Vert19}
D.~Vert, R.~Sirdey, and S.~Louise, ``On the limitations of the chimera graph
  topology in using analog quantum computers,'' in \emph{Proceedings of the
  16th ACM International Conference on Computing Frontiers}.\hskip 1em plus
  0.5em minus 0.4em\relax Proceedings of the 16th ACM International Conference
  on Computing Frontiers, Apr. 2019, pp. 226–--229.

\bibitem{Advantage}
C.~McGeoch and P.~Farr\'{e}, ``The {D-Wave} {Advantage} {System}: An
  overview,'' D-Wave Systems, Tech. Rep., 2020.

\bibitem{DWaveAdvantage19}
K.~Boothby, P.~Bunyk, J.~Raymond, and A.~Roy, ``Next-generation topology of
  {D-Wave} quantum processors,'' D-Wave Systems, techreport 14-1026A-C, Feb.
  2019.

\bibitem{Tanahashi19}
\BIBentryALTinterwordspacing
K.~Tanahashi, S.~Takayanagi, T.~Motohashi, and S.~Tanaka, ``Application of
  {Ising} machines and a software development for {Ising} machines,''
  \emph{Journal of the Physical Society of Japan}, vol.~88, no.~6, p. 061010,
  2019. [Online]. Available: \url{https://doi.org/10.7566/JPSJ.88.061010}
\BIBentrySTDinterwordspacing

\bibitem{Tao20}
M.~Tao, K.~Nakano, Y.~Ito, R.~Yasudo, M.~Tatekawa, R.~Katsuki, T.~Yazane, and
  Y.~Inaba, ``A work-time optimal parallel exhaustive search algorithm for the
  {QUBO} and the {Ising} model, with {GPU} implementation,''
  \emph{International Parallel and Distributed Processing Symposium Workshops},
  pp. 557--566, May 2020.

\bibitem{Imanaga21}
T.~Imanaga, K.~Nakano, R.~Yasudo, Y.~Ito, Y.~Kawamata, R.~Katsuki, S.~Ozaki,
  T.~Yazane, and K.~Hamano, ``Solving the sparse {QUBO} on multiple {GPUs} for
  simulating a quantum annealer,'' in \emph{Proc. of International Symposium on
  Computing and Networking}, Nov. 2021, pp. 19--28.

\bibitem{Zaborniak21}
T.~Zaborniak and R.~de~Sousa, ``Benchmarking {Hamiltonian} noise in the
  {D-Wave} quantum annealer,'' \emph{IEEE Transactions on Quantum Engineering},
  Jan. 2021.

\bibitem{Oku19}
D.~Oku, K.~Terada, M.~Hayashi, M.~Yamaoka, S.~Tanaka, and N.~Togawa, ``A
  fully-connected {Ising} model embedding method and its evaluation for {CMOS}
  annealing machines,'' \emph{{IEICE} Trans. Inf. Syst.}, vol. 102-D, no.~9,
  pp. 1696--1706, 2019.

\bibitem{Yamamoto21}
K.~Yamamoto, K.~Kawamura, K.~Ando, N.~Mertig, T.~Takemoto, M.~Yamaoka,
  H.~Teramoto, A.~Sakai, S.~Takamaeda-Yamazaki, and M.~Motomura, ``{STATICA}: A
  512-spin 0.25m-weight annealing processor with an all-spin-updates-at-once
  architecture for combinatorial optimization with complete spin–spin
  interactions,'' \emph{IEEE Journal of Solid-State Circuits}, vol.~56, no.~1,
  pp. 165--178, 2021.

\bibitem{Kagawa21}
H.~Kagawa, Y.~Ito, K.~Nakano, R.~Yasudo, Y.~Kawamata, R.~Katsuki, Y.~Tabata,
  T.~Yazane, and K.~Hamano, ``High-throughput {FPGA} implementation for
  quadratic unconstrained binary optimization,'' \emph{Concurrency and
  Computation: Practice and Experience}, p. e6565, Aug. 2021.

\bibitem{Goto21}
H.~Goto, K.~Endo, M.~Suzuki, Y.~Sakai, T.~Kanao, Y.~Hamakawa, R.~Hidaka,
  M.~Yamasaki, and K.~Tatsumura, ``High-performance combinatorial optimization
  based on classical mechanics,'' \emph{Science Advances}, vol.~7, no.~6, Feb.
  2021.

\bibitem{Okuyama19}
T.~Okuyama, T.~Sonobe, K.~Kawarabayashi, and M.~Yamaoka, ``Binary optimization
  by momentum annealing,'' \emph{PHYSICAL REVIEW E}, vol. 100, no.~1, p.
  012111, Jul. 2019.

\bibitem{Yasudo-JPDC22}
R.~Yasudo, K.~Nakano, Y.~Ito, R.~Katsuki, Y.~Tabata, T.~Yazane, and K.~Hamano,
  ``{GPU}-accelerated scalable solver with bit permutated cyclic-min algorithm
  for quadratic unconstrained binary optimization,'' \emph{Journal of Parallel
  and Distributed Computing}, vol. 167, pp. 109--122, Sep. 2022.

\bibitem{Inagaki16}
T.~Inagaki, Y.~Haribara, K.~Igarashi, T.~Sonobe, S.~Tamate, T.~Honjo,
  A.~Marandi, P.~L. McMahon, T.~Umeki, K.~Enbutsu \emph{et~al.}, ``A coherent
  {Ising} machine for 2000-node optimization problems,'' \emph{Science}, vol.
  354, no. 6312, pp. 603--606, 2016.

\bibitem{Honjo21}
T.~Honjo, T.~Sonobe, K.~Inaba, T.~Inagaki, T.~Ikuta, Y.~Yamada, T.~Kazama,
  K.~Enbutsu, T.~Umeki, R.~Kasahara, K.~Kawarabayashi, and H.~Takesue,
  ``100,000-spin coherent {Ising} machine,'' \emph{Science Advances}, vol.~7,
  no.~40, p. eabh0952, 2021.

\bibitem{D-Wave-Hybrid20}
C.~McGeoch, P.~Farr\'{e}, and W.~Bernoudy, ``{D-Wave} hybrid solver service +
  {Advantage}: Technology update,'' D-Wave Systems, Tech. Rep., 2020.

\bibitem{nofreelunch17}
D.~H. Wolpert and W.~G. Macready, ``No free lunch theorems for optimization,''
  \emph{{IEEE} Trans. Evol. Comput.}, vol.~1, no.~1, pp. 67--82, 1997.

\bibitem{Whitley99}
D.~Whitley, S.~Rana, and R.~B. Heckendorn, ``The island model genetic
  algorithm: On separability, population size and convergence,'' \emph{Journal
  of Computing and Information Technology}, vol.~7, no.~1, pp. 3--47, 1999.

\bibitem{Gurobi}
``Gurobi optimizer reference manual,''
  https://www.gurobi.com/documentation/9.0/refman/index.html.

\bibitem{Glover18}
F.~W. Glover and G.~A. Kochenberger, ``A tutorial on formulating {QUBO}
  models,'' \emph{CoRR}, vol. abs/1811.11538, 2018.

\bibitem{Tosun22}
U.~Tosun, ``A new tool for automated transformation of {Q}uadratic {A}ssignment
  {P}roblem instances to quadratic unconstrained binary optimisation models,''
  \emph{Expert Systems with Applications}, Sep. 2022.

\bibitem{Laarhoven87}
P.~J.~M. van Laarhoven and E.~H.~L. Aarts, \emph{Simulated Annealing: Theory
  and Applications (Mathematics and Its Applications Book 37)}.\hskip 1em plus
  0.5em minus 0.4em\relax Springer, 1987.

\bibitem{Glover97}
F.~Glover and M.~Laguna, ``Tabu search,'' in \emph{Handbook of Combinatorial
  Optimization}, D.-Z. Du and P.~M. Pardalos, Eds.\hskip 1em plus 0.5em minus
  0.4em\relax Kluwer Academic Publishers, 1997, pp. 2093--–2229.

\bibitem{Whitley98}
D.~Whitley, S.~Rana, and R.~B. Heckendorn, ``The island model genetic
  algorithm: On separability, population size and convergence,'' \emph{Journal
  of Computing and Information Technology}, vol.~7, pp. 33--47, 1998.

\bibitem{CUDA-Programming}
{NVIDIA Corporation}, ``{NVIDIA} {CUDA} {C} programming guide version 11.2.2,''
  https://docs.nvidia.com/cuda/cuda-c-programming-guide/index.html, Nov. 2021.

\bibitem{NVIDIA-A100}
------, ``{NVIDIA} {A100} tensor core {GPU} architecture,''
  https://images.nvidia.com/aem-dam/en-zz/Solutions/data-center/nvidia-ampere-architecture-whitepaper.pdf,
  2020.

\bibitem{Matsumoto98}
M.~Matsumoto and T.~Nishimura, ``{Mersenne} twister: a 623-dimensionally
  equidistributed uniform pseudo-random number generator,'' \emph{ACM
  Transactions on Modeling and Computer Simulation}, vol.~8, no.~1, pp. 3--30,
  Jan. 1998.

\bibitem{Marsaglia03}
G.~Marsaglia, ``Xorshift {RNGs},'' \emph{Journal of Statistical Software},
  vol.~8, Jul. 2003.

\bibitem{ABCI2.0}
S.~Takizawa, Y.~Tanimura, H.~Nakada, R.~Takano, and H.~Ogawa, ``{ABCI} 2.0:
  Advances in open {AI} computing infrastructure at {AIST},'' IPSJ SIG,
  techreport HPC-180, Jul. 2021.

\bibitem{k2000}
S.~Tamate, T.~Sonobe, and Y.~Haribara, ``Simulated annealing for max-cut
  problems on \{+1,-1\}-weighted complete graphs,''
  \url{https://github.com/hariby/SA-complete-graph}, 2016.

\bibitem{gset}
Y.~Ye, ``http://www.stanford.edu/yyye/yyye/gset.''

\bibitem{Goto19}
H.~Goto, K.~Tatsumura, and A.~R. Dixon, ``Combinatorial optimization by
  simulating adiabatic bifurcations in nonlinear hamiltonian systems,''
  \emph{Science Advances}, vol.~5, no.~4, Apr. 2019.

\bibitem{Burkard91}
R.~E. Burkard, S.~Karisch, and F.~Rendl, ``{QAPLIB}-a quadratic assignment
  problem library,'' \emph{European Journal of Operational Research}, vol.~55,
  no.~1, pp. 115--119, Nov. 1991.

\bibitem{Atobe21}
Y.~Atobe, M.~Tawada, and N.~Togawa, ``Hybrid annealing method based on
  {subQUBO} model extraction with multiple solution instances,'' \emph{IEEE
  Transactions on Computers}, 2021.

\end{thebibliography}

\end{document}